\documentclass[conference]{IEEEtran}
\IEEEoverridecommandlockouts

\usepackage{cite}
\usepackage{amsmath,amssymb,amsfonts}
\usepackage{graphicx}
\usepackage{textcomp}
\usepackage{xcolor}
\usepackage{array}
\usepackage{booktabs}
\usepackage{tabularx}
\usepackage{acro}

\newcommand{\bt}[1]{#1}

\newcommand{\ket}[1]{\lvert #1\rangle}

\acsetup{first-style=long-short}

\DeclareAcronym{rf}{
  short = {RF},
  long = {radio frequency}
}
\DeclareAcronym{doa}{
  short = {DoA},
  long = {direction-of-arrival}
}
\DeclareAcronym{lo}{
  short = {LO},
  long = {local oscillator}
}
\DeclareAcronym{rhcp}{
  short = {RHCP},
  long = {right-hand circularly polarised}
}
\DeclareAcronym{eit}{
  short = {EIT},
  long = {electromagnetically induced transparency}
}
\DeclareAcronym{isac}{
  short = {ISAC},
  long = {integrated sensing and communication}
}
\DeclareAcronym{mimo}{
  short = {MIMO},
  long = {multiple-input multiple-output}
}
\DeclareAcronym{fft}{
  short = {FFT},
  long = {fast Fourier transform}
}
\DeclareAcronym{qsn}{
  short = {QSN},
  long = {quantum shot noise}
}
\DeclareAcronym{va}{
  short = {VA},
  long = {vector antenna}
}
\DeclareAcronym{rmse}{
  short = {RMSE},
  long = {root-mean-square error}
}
\DeclareAcronym{cdf}{
  short = {CDF},
  long = {cumulative distribution function}
}

\def\BibTeX{{\rm B\kern-.05em{\sc i\kern-.025em b}\kern-.08em
    T\kern-.1667em\lower.7ex\hbox{E}\kern-.125emX}}
\begin{document}
\bstctlcite{IEEEexample:BSTcontrol}

\title{{A Three-Axis Rydberg Atomic Receiver Model for Polarization-Based Direction-of-Arrival Sensing}}

\author{

\IEEEauthorblockN{Bangyan Xia, Bo Tan}
\IEEEauthorblockA{Department of Electronics and Electrical Engineering, University College London\\
London, United Kingdom, \\ \{bangyan.xia.24, tan.bo\}@ucl.ac.uk}
}

\maketitle

\begin{abstract}
This paper develops a three-axis Rydberg atomic receiver model for \ac{doa} estimation of the \ac{rf} signal source. Three nominally orthogonal \ac{rf} \acp{lo}, each assigned a distinct frequency offset, project the weak incident \ac{rf} field into resolvable complex beat coefficients in the optical readout. A beat-vector observation is obtained by adopting a response calibration mechanism that maps the coefficients onto the three axes of the Cartesian coordinate system. For a known circularly polarised \ac{rf} signal, this observation induces a direction-dependent Rydberg steering vector, enabling maximum-likelihood \ac{doa} estimation. The proposed procedure translates the Rydberg beat-vector front end into a highly sensitive \ac{rf} field \ac{doa} sensor. Simulation results show that trace-level beat extraction supports the proposed \ac{doa} sensing workflow and provides a basis for comparison with a traditional \ac{va} benchmark.
\end{abstract}

\begin{IEEEkeywords}
Rydberg atomic receiver, beat-vector readout, direction-of-arrival estimation, vector sensing.
\end{IEEEkeywords}

\acresetall

\section{Introduction}
\label{sec:introduction}

Future wireless systems are expected to operate over an increasingly wide and reconfigurable spectrum for combining communication, sensing, and localisation. Such systems require frequency-agile \ac{rf} front-ends that preserve field amplitude, phase, polarisation, and direction information. Conventional metallic antennas and arrays remain highly effective, but their responses depend on geometry, matching networks, mutual coupling, and frequency-dependent calibration. These issues become more restrictive when wideband operation, small apertures, and dense \ac{isac} nodes are needed. Rydberg atomic receivers offer an alternative route by exploiting atom–field interactions rather than current induction in metallic structures. {In a typical implementation, probe and coupling lasers excite alkali atoms to Rydberg states, where large electric-dipole moments enable strong coupling between incident \ac{rf} fields and adjacent Rydberg levels. The resulting perturbation of the \ac{eit} response, manifested as Autler–Townes splitting or Stark shifts, enables readout of the \ac{rf} field via a photodiode.} As the transition frequencies are determined by selectable atomic energy levels, this atom-referenced mechanism enables a sensitive and frequency-reconfigurable front end for communication and sensing~\cite{anderson2020rfcomms,gong2025raqwireless}.

Recent works have examined Rydberg receivers as \ac{rf} front-ends for both communication and sensing applications. System-level studies have integrated such atomic receivers into classical wireless communication architectures~\cite{chen2025harnessing} and theoretical models~\cite{cui2025frontier}. Analyses and demonstrations have also been reported for multi-band sensing~\cite{kim2025multiband}, high-order modulation~\cite{zhu2025qam}, and \ac{isac}~\cite{chen2025isac}. Rydberg receivers have further been investigated in \ac{mimo} architectures for multi-user communications~\cite{cui2025atomicmimo,gong2025mimouplink}, as well as in spatial-processing configurations for \ac{doa} estimation~\cite{guo2025aoa,gong2025multitargetdoa}. 
Within the broader communication-and-sensing context, Rydberg \ac{doa} estimation has been realised through several different physical mechanisms. Array-based architectures obtain spatial diversity from multiple atomic sensing elements~\cite{gong2025multitargetdoa}, whereas recent single-cell studies have shown that angular information can also be recovered through inner-vapor interference~\cite{guo2025aoa} or polarisation-aware field measurements~\cite{chen2025polarizationdoa}. Separately, Rydberg vector electrometry has demonstrated sensitivity to \ac{rf} fields~\cite{sedlacek2013vector}, and complete three-dimensional polarimetry has shown that orthogonal, frequency-separated \acp{lo} can be used to reconstruct and calibrate Cartesian field components~\cite{elgee2024vectorpolarimetry}.
%
Prior work has demonstrated single-cell Rydberg direction finding~\cite{elgee2025kvector}. Based on ~\cite{elgee2025kvector}, we developed a consistent statistical model that connects calibrated beat measurements to maximum-likelihood estimation for higher \ac{doa}.

In this work, we develop a three-axis Rydberg beat-vector receiver model for polarisation-based \ac{doa} sensing. In \bt{the adopted architecture}, three nominally orthogonal \ac{rf} \acp{lo} are assigned distinct frequency offsets, allowing the Cartesian projections of a weak incident \ac{rf} field to be encoded into separable complex beat coefficients within a scalar optical readout. Response normalisation then maps the extracted coefficients to a calibrated three-axis field observation. For a known \ac{rhcp} \ac{rf} sensing waveform \bt{under a single-plane-wave assumption}, this observation admits a direction-dependent steering-vector representation and enables \bt{covariance-weighted maximum-likelihood \ac{doa} estimation with unknown complex field amplitude. The model also provides a common incident-field basis for comparison with a conventional \ac{va}.}


\section{System Model: Three-{axis} Rydberg Beat-Vector Readout}
\label{sec:three_frequency_readout}

\begin{figure}[t]
\centering
\includegraphics[width=0.8\linewidth]{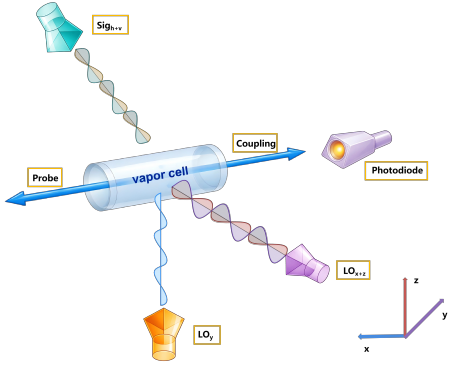}
\caption{\textbf{Single-cell three-axis Rydberg receiver.} Three orthogonally polarised, frequency-offset \acp{lo} define the laboratory axes. Counter-propagating probe and coupling beams provide the optical readout, whose heterodyne beat tones form the three-axis observation.}
\label{fig:system_architecture}
\end{figure}

\bt{Fig.~\ref{fig:system_architecture}} summarises the system architecture modelled
in this section. Three nominally orthogonal, frequency-offset \ac{rf} \acp{lo} are applied to a single Rydberg vapor cell. A photodiode is used to read the probe laser signal, which can be {digitised} to obtain the corresponding beat tones and then normalised into the vector observation used for the \ac{doa} estimation.

\subsection{Rydberg Atomic Receiver Preliminaries}
\label{subsec:rydberg_preliminaries_conf}
A Rydberg atomic receiver maps an incident \ac{rf} electric field to an optical observable by exploiting the field-sensitive response of highly excited atomic states. We consider the standard vapor-cell configuration, where a weak probe laser and a coupling laser excite alkali atoms through a ladder-type \ac{eit} scheme. At the level needed for the \ac{doa} sensing model, we use compact state labels and represent the receiver by the four-level ladder $\ket{1}
\overset{\mathrm{probe},\,\Omega_{\mathrm p}}{\longleftrightarrow}
\ket{2}
\overset{\mathrm{coupling},\,\Omega_{\mathrm c}}{\longleftrightarrow}
\ket{3}
\overset{\mathrm{RF},\,\Omega_{\mathrm{RF}}}{\longleftrightarrow}
\ket{4}.$
Here $\ket{1}$ is a low-lying atomic state, $\ket{2}$ is an intermediate optically excited state, and $\ket{3}$ and $\ket{4}$ label the neighboring Rydberg manifolds coupled by the \ac{rf} field. The probe and coupling fields form the optical readout on the $\ket{1}$--$\ket{2}$--$\ket{3}$ ladder, while the incident \ac{rf} field drives the Rydberg transition $\ket{3}\leftrightarrow\ket{4}$, thereby perturbing the \ac{eit} response observed from the transmitted probe beam.

{For a local complex \ac{rf}-field envelope $\tilde{\mathbf{E}}$, the \ac{rf} coupling strength is characterised by the Rabi frequency $\Omega_{\mathrm{RF}}$:}
\begin{equation}
    \Omega_{\mathrm{RF}} = \frac{\mathbf{d}_{34} \cdot \tilde{\mathbf{E}}}{\hbar},
    \label{eq:rf_rabi_frequency}
\end{equation}
{where $\mathbf d_{34}$ is the electric-dipole matrix element of the $\ket{3}\leftrightarrow\ket{4}$ transition and the factor $1/\hbar$ converts the dipole-interaction energy to an angular Rabi frequency. This coupling modifies the \ac{eit} spectrum and, on resonance, produces an Autler--Townes splitting of the probe laser transmission that tracks $|\Omega_{\mathrm{RF}}|$. Reading this optical feature alone realises scalar \ac{rf} electrometry: it measures the magnitude of the field projection onto a single dipole orientation, but retains neither the carrier phase nor the orientation of the incident field, both of which are required for \ac{doa} sensing.}

{The phase is recovered by heterodyne operation. A strong \ac{rf} \ac{lo}, detuned from the signal carrier by a small offset, is applied to the same Rydberg vapor cell. The weak incident field {interferes} coherently with the \ac{lo}, so the optical readout exhibits a slow modulation at the offset frequency whose amplitude and phase reproduce those of the incident-field projection relative to the \ac{lo} reference~\cite{sedlacek2013vector}. A single \ac{lo}, however, still reads out only one polarisation projection of the field.}

{The direction of the incident \ac{rf} field is recovered from the vector nature of the atomic response itself. Since $\ket{3}$ and $\ket{4}$ are angular-momentum manifolds containing several magnetic sublevels rather than isolated two-level states, each polarisation component of the local field drives its own set of sublevel transitions, with relative strengths fixed by the Clebsch--Gordan coefficients of the transition; for a transition whose total angular momentum increases from $J$ to $J+1$, every sublevel couples to all three polarisation components, so a single vapor cell responds to the complete vector field~\cite{elgee2024vectorpolarimetry}.
Three orthogonally polarised \acp{lo} define the laboratory readout axes $(x,y,z)$. Each \ac{lo} selects the field projection along its corresponding polarisation axis. For axis $i$, the fixed Clebsch--Gordan weighting is absorbed into a calibrated Rabi-per-field coefficient $\kappa_i$, where $i\in\{x,y,z\}$. This coefficient relates the field projection to its effective Rabi contribution.
Assigning a distinct frequency offset to each \ac{lo} then separates the three projections into distinct beat tones on the same optical trace, which is the three-axis beat extraction developed in the next subsection.}

\subsection{Three-{axis} Beat Extraction}
\label{subsec:three_frequency_extraction_conf}

{All three readout channels are heterodyne measurements on the same \ac{rf}-coupled Rydberg transition, distinguished only by their \ac{lo} frequency offsets; they do not represent three independent Cartesian atomic transitions.} Let the three \ac{lo} components be $\mathbf E_{\mathrm{LO}}(t)=[E_{\mathrm{LO},x}(t),E_{\mathrm{LO},y}(t), E_{\mathrm{LO},z}(t)]^T$, with $E_{\mathrm{LO},i}(t)=A_{\mathrm{LO},i} \cos((\omega-\Delta\omega_i)t+\psi_{\mathrm{LO},i})$ for $i\in\{x,y,z\}$. Here $\omega=2\pi f_0$ is the \ac{rf} carrier angular frequency, $\Delta\omega_i$ is the signed offset of the signal carrier relative to the \ac{lo} on axis $i$, and $\psi_{\mathrm{LO},i}$ is the known \ac{lo} phase reference. The weak incident \ac{rf} sensing field is represented by the complex phasor $\tilde{\mathbf E}_{\mathrm s}=[\tilde E_x\ \tilde E_y\ \tilde E_z]^T$, with $\tilde E_i=A_i e^{j\psi_i}$ at the receiver calibration plane; the corresponding passband \ac{rf} field is $\mathbf E_{\mathrm s}(t)=\Re\{\tilde{\mathbf E}_{\mathrm s}e^{j\omega t}\}$.

\bt{We use $\Delta\omega_i=2\pi\Delta f_i$, where $\Delta f_i$ is the signed beat frequency in hertz.} \bt{The three \ac{rf} \acp{lo} share a common coherent reference. Their phase offsets relative to the digital demodulation references are assumed known, compensated, and stable over the coherent integration window $T_{\mathrm{int}}$ used for beat extraction. The unknown common signal phase is absorbed into $\alpha$ below.}

Under strong \ac{lo} and weak incident signal conditions, $A_i\ll A_{\mathrm{LO},i}$, the atomic and optical responses can be {expanded to first order around the \ac{lo}-only operating point}. The key \ac{rf}-domain step is the coherent addition of the strong \ac{lo} and the weak incident field on each polarisation axis. After removing the common carrier at $\omega$, the slowly varying envelope on axis $i$ is $\bar E_i(t)=A_{\mathrm{LO},i}
e^{j(-\Delta\omega_i t+\psi_{\mathrm{LO},i})}+A_i e^{j\psi_i}$. The \ac{rf} amplitude that perturbs the Rydberg transition on that axis is $q_i(t)\triangleq |\bar E_i(t)|$, namely
\begin{equation}
q_i(t) = \left[ A_{\mathrm{LO},i}^2 + A_i^2 + 2A_{\mathrm{LO},i}A_i
\cos(\Delta\omega_i t+\Delta\psi_i) \right]^{1/2},
\label{eq:axis_exact_envelope_magnitude_conf}
\end{equation}
where $\Delta\psi_i=\psi_i-\psi_{\mathrm{LO},i}$ is the signal phase relative to the \ac{lo} reference. Since $A_i\ll A_{\mathrm{LO},i}$, a first-order expansion of
\eqref{eq:axis_exact_envelope_magnitude_conf} gives
\begin{equation}
q_i(t) \approx A_{\mathrm{LO},i} + A_i\cos(\Delta\omega_i t+\Delta\psi_i).
\label{eq:axis_small_signal_conf}
\end{equation}
This expression is the heterodyne origin of the beat tone: the \ac{lo} provides the
phase reference, while the weak \ac{rf}-field projection appears as a small amplitude
modulation at $\Delta\omega_i$.

{The balanced effective three-axis Rabi approximation model can be used after calibration.} The approximation assumes that the three \ac{lo} polarisations define an orthogonal local basis, and that their intrinsic polarisation gains have been equalised by \ac{lo} power setting and response calibration~\cite{elgee2024vectorpolarimetry}. With
$\Omega_i(t)=\kappa_i q_i(t)$ for $i\in\{x,y,z\}$, the \ac{rf} feature that controls
the local \ac{eit}/Autler--Townes response is represented by the Euclidean norm of
the three effective Rabi components,
\begin{equation}
\Omega_{\mathrm{eff}}(t) \approx
\left(\Omega_x^2(t) + \Omega_y^2(t) + \Omega_z^2(t) \right)^{1/2}.
\label{eq:rabi_vector_magnitude_conf}
\end{equation}

{This quadrature form holds for a polarisation-symmetric response, represented here by an idealised uniform magnetic-sublevel population and an effective Clebsch--Gordan-weighted dipole.} In the balanced reference case, $\kappa_x\simeq\kappa_y\simeq\kappa_z$ after \ac{lo} power equalization and receiver calibration.

Let the \ac{lo}-only Rabi magnitude over the coherent processing window be
$\Omega_0=[(\kappa_xA_{\mathrm{LO},x})^{2}
+(\kappa_yA_{\mathrm{LO},y})^{2}
+(\kappa_zA_{\mathrm{LO},z})^{2}]^{1/2}$. Substituting
\eqref{eq:axis_small_signal_conf} and the per-axis Rabi definition into \eqref{eq:rabi_vector_magnitude_conf}, and retaining only first-order weak-field terms, gives a baseline-plus-modulation form:
\begin{equation}
\Omega_{\mathrm{eff}}(t) \approx \Omega_0 + \sum_{i\in\{x,y,z\}} \beta_i A_i\cos(\Delta\omega_i t+\Delta\psi_i),
\label{eq:rabi_norm_linearization_conf}
\end{equation}
where $\beta_i=\kappa_i^2A_{\mathrm{LO},i}/\Omega_0$. The first term in \eqref{eq:rabi_norm_linearization_conf} is the \ac{lo}-only contribution. The remaining terms are the three first-order beat modulations {caused by the difference between the \ac{lo} and the incident \ac{rf} field}, which are weighted by the \ac{lo} operating point. The distinct offsets $\Delta\omega_i$ assign {the projections on three Cartesian axes} to different beat bins.

The measured optical trace is denoted by $y(t)$. It corresponds to a probe-transmission sample, a fitted \ac{eit} feature, or another scalar optical observable. Linearising the selected optical feature with respect to $\Omega_{\mathrm{eff}}$ around the \ac{lo}-only point maps the beat modulations directly into the scalar trace. For an ideal axis-aligned receiver,
\begin{equation}
y(t)\approx y_0(t) + \sum_{i\in\{x,y,z\}}\gamma_i A_i\cos(\Delta\omega_i t+\Delta\psi_i) + n(t),
\label{eq:ideal_optical_beat_conf}
\end{equation}
where {$\gamma_i=s_{\Omega}\beta_i$}, $y_0(t)$ includes the \ac{lo}-only optical response associated with $\Omega_0$, and $s_{\Omega}$ is the local optical-feature slope. Equivalently, since $\gamma_iA_i\cos(\Delta\omega_i t+\Delta\psi_i)=\Re\{\gamma_i e^{-j\psi_{\mathrm{LO},i}}\tilde E_i\,e^{j\Delta\omega_i t}\}$, coherent demodulation at the signed offset $\Delta\omega_i$ followed by removal of the known \ac{lo} phase $\psi_{\mathrm{LO},i}$ yields the ideal complex beat coefficient $\gamma_i\tilde E_i$.

In a practical receiver, \ac{rf}-field coupling to the vapor cell, atomic selection rules, finite \ac{lo} orthogonality, vapor-cell scattering, and optical feature fitting can mix the Cartesian components before they appear in a beat bin. With the known \ac{lo} phases compensated, the ideal coefficient $\gamma_i\tilde E_i$ is therefore replaced by a general linear sensitivity $\mathbf c_i^H\tilde{\mathbf E}_{\mathrm s}$. Under this local first-order model, the trace can be written as a slowly varying baseline plus three narrowband beat terms,
\begin{equation}
\begin{aligned}
y(t) &\approx y_0(t) + \Re\left\{ \sum_{i\in\{x,y,z\}} \left(\mathbf c_i^{H}\tilde{\mathbf E}_{\mathrm s}\right) e^{j\Delta\omega_i t} \right\} + n(t),
\end{aligned}
\label{eq:trace_linearization_conf}
\end{equation}
where $\mathbf c_i^H$ is the complex local sensitivity of the $i$th beat bin to the three Cartesian \ac{rf}-field components and $n(t)$ is optical-readout and feature-estimation noise. The baseline $y_0(t)$ is the \ac{lo}-only optical term. In the ideal model it is constant over the coherent processing window, and therefore appears only at the DC bin. Thus the uncalibrated complex coefficient of the $i$th beat tone is obtained by coherent Fourier or lock-in processing of the measured trace:
\begin{equation}
\xi_i = \frac{2}{T_{\mathrm{int}}} \int_{t_0}^{t_0+T_{\mathrm{int}}} y(t) e^{-j\Delta\omega_i t}\,dt, \quad i\in\{x,y,z\}.
\label{eq:raw_beat_extraction_conf}
\end{equation}
where $t_0$ is the start time of the integration window. This extraction assumes that the three frequency offsets are resolvable over the integration window and each lock-in reference remains phase coherent over $T_{\mathrm{int}}$. Since $y(t)$ is real, each tone also has a conjugate image
at $-\Delta\omega_i$. The chosen offset avoids overlap between any desired beat bin and another tone or conjugate image, i.e., $\Delta\omega_i\neq\pm\Delta\omega_j$ for $i\neq j$. In a sampled implementation, the same operation is performed by reading the corresponding signed \ac{fft} bins. \bt{Thus, the $-200$~kHz $z$-axis value reported below is the signed complex-baseband offset for an \ac{lo} placed $200$~kHz above the carrier; the $+200$~kHz component is its conjugate image.} Stacking the three extracted coefficients gives the raw beat vector $\boldsymbol\xi=[\xi_x,\xi_y,\xi_z]^T$, which converts a scalar optical trace into a complex three-axis \ac{rf} observation.

\subsection{Response Normalisation and Noise Model}
\label{subsec:response_normalization_noise_model}

\bt{Stacking the sensitivities in \eqref{eq:trace_linearization_conf} gives $\boldsymbol\xi=\mathbf C{\mathrm R}\tilde{\mathbf E}{\mathrm s}+\boldsymbol\eta$, where the rows of $\mathbf C_{\mathrm R}$ are $\mathbf c_i^H$. After \ac{lo} amplitude/phase calibration and correction of stray \ac{rf} reflections, which account for the residual cross-axis terms, $\mathbf C_{\mathrm R}$ reduces to a scaled identity~\cite{elgee2024vectorpolarimetry}; we therefore take $\mathbf C_{\mathrm R}=c_{\mathrm R}\mathbf I_3$, yielding}
\begin{equation}
\boldsymbol\xi = c_{\mathrm R}\tilde{\mathbf E}_{\mathrm s} + \boldsymbol\eta,
\label{eq:raw_response_model_conf}
\end{equation}
where $c_{\mathrm R}$ is the common complex readout gain after balanced $x$-, $y$-, and $z$-axis \ac{lo} setting, and $\boldsymbol\eta$ is the post-extraction noise. A {surrogate} scalar estimate $\widehat c_{\mathrm R}$ is obtained offline from a known calibration field, and the normalized Rydberg sensing vector is $\mathbf b_{\mathrm R}\triangleq \widehat c_{\mathrm R}^{-1}\boldsymbol\xi$. Substituting \eqref{eq:raw_response_model_conf} gives
\begin{equation}
\mathbf b_{\mathrm R} = \mathbf A_{\mathrm R}\tilde{\mathbf E}_{\mathrm s} + \mathbf w_{\mathrm R},
\quad
\mathbf A_{\mathrm R} \triangleq \frac{c_{\mathrm R}}{\widehat c_{\mathrm R}}\mathbf I_3,
\quad
\mathbf w_{\mathrm R} \triangleq \widehat c_{\mathrm R}^{-1}\boldsymbol\eta.
\label{eq:normalized_rydberg_observation_conf}
\end{equation}
The numerical results use the idealised response-normalised case $\widehat c_{\mathrm R}=c_{\mathrm R}$, so that $\mathbf A_{\mathrm R}=\mathbf I_3$. \bt{For a general nonsingular response, matrix calibration~\cite{elgee2024vectorpolarimetry} gives $\mathbf b_{\mathrm R}=\widehat{\mathbf C}_{\mathrm R}^{-1}\boldsymbol\xi$ and noise covariance $\widehat{\mathbf C}_{\mathrm R}^{-1}\mathbf R_{\eta}(\widehat{\mathbf C}_{\mathrm R}^{-1})^H$, where $\mathbf R_{\eta}$ is the covariance of $\boldsymbol\eta$.}

The post-extraction noise is separated into two physical terms: external and internal noise. External noise is a pre-receiver \ac{rf} field perturbation $\mathbf e_{\mathrm{ext}}$, while internal noise $\boldsymbol\eta_{\mathrm{int}}$ collects atomic \ac{qsn}, optical readout noise, detector noise, and feature-estimation error. {With $\mathbf R_{\mathrm{ext}}$ and $\mathbf R_{\mathrm{int}}$ denoting the covariances of $\mathbf e_{\mathrm{ext}}$ and $\boldsymbol\eta_{\mathrm{int}}$, the normalised covariance of $\boldsymbol\eta=c_{\mathrm R}\mathbf e_{\mathrm{ext}}+ \boldsymbol\eta_{\mathrm{int}}$ is $\mathbf R_{\mathrm R} = |c_{\mathrm R}/\widehat c_{\mathrm R}|^2\mathbf R_{\mathrm{ext}} + |\widehat c_{\mathrm R}|^{-2}\mathbf R_{\mathrm{int}}$.}

{The internal atomic contribution is assigned a \ac{qsn}-limited floor. The equivalent minimum detectable field is
\begin{equation}
E_{\min}=h/(\|\mathbf d_{34}\|\sqrt{T_iT_rN_a}),\label{eq:qsn_field_floor_conf}
\end{equation}
where $h=2\pi\hbar$, $T_i$ is the atomic measurement time (set equal to the Fourier integration window, $T_i=T_{\mathrm{int}}=T_{\mathrm{obs}}$), $T_r$ is the Rydberg relaxation time, $N_a$ is the participating atom number, and $\|\mathbf d_{34}\|$ is the effective \ac{rf} transition dipole norm.} In the baseline numerical model we use response-normalised, field-domain beat coordinates and approximate the total covariance as isotropic~\cite{cui2025atomicmimo},
\begin{equation}
\mathbf w_{\mathrm R}\sim\mathcal{CN}(\mathbf 0,\mathbf R_{\mathrm R}),
\qquad
\mathbf R_{\mathrm R} =\left( \sigma_{\mathrm{int}}^2 +
\sigma_{\mathrm{ext}}^2 \right)\mathbf I_3 .
\label{eq:rydberg_noise_conf}
\end{equation}
Here $\sigma_{\mathrm{int}}^2=E_{\min}^2$ is the per-axis internal field-noise variance, taken as the \ac{qsn} floor, and $\sigma_{\mathrm{ext}}^2$ is the per-axis external field-noise variance, i.e., $\mathbf R_{\mathrm{ext}}=\sigma_{\mathrm{ext}}^2\mathbf I_3$~\cite{cui2025atomicmimo}. All field phasors use the peak convention, and each complex noise variance denotes $\mathbb E[|w_i|^2]$ in $(\mathrm{V/m})^2$. The \ac{doa} sensing model below uses the response-normalised receiver, for which $\mathbf A_{\mathrm R}\approx\mathbf I_3$. The known \ac{rf} sensing waveform then produces a normalised observation denoted $\mathbf b_{\mathrm s}$, which is the specialisation of $\mathbf b_{\mathrm R}$ to that waveform, and we write $\mathbf R_w=\mathbf R_{\mathrm R}$ {as} its beat-domain noise covariance.

\section{{Direction-of-arrival} Estimation}
\label{sec:known_waveform_doa}

{The three-axis beat-vector readout in Section~\ref{sec:three_frequency_readout} preserves the polarisation-dependent Cartesian projection pattern of the incident \ac{rf} field, which varies with propagation direction for a known circularly polarised sensing waveform. We therefore estimate the \ac{doa} using the angle pair $(\theta,\phi)$, where $\theta$ is the polar angle from the positive $z$ axis and $\phi$ is the azimuth in the $x$-$y$ plane.}

\subsection{\acs{rf} Field Steering Model}
\label{subsec:rhcp_sensing_steering}

{Conventional three-dimensional \acp{va} have been studied as compact direction-finding sensors and experimentally validated for angle-of-arrival estimation~\cite{obaid2023using}. We adopt a similar principle for \ac{doa} based on the Rydberg beat-vector. Here $\hat{\mathbf k}(\theta,\phi)$ denotes the unit \ac{doa} vector. The vectors $\hat{\boldsymbol\theta}(\theta,\phi)$ and $\hat{\boldsymbol\phi}(\phi)$ form the local transverse basis orthogonal to $\hat{\mathbf k}$, and $\mathbf U(\theta,\phi)$ maps a Jones vector in that transverse basis to Cartesian field components~\cite{varshney2025tripole}:}
\begin{equation}
\begin{aligned}
\hat{\mathbf k}(\theta,\phi) &=
\begin{bmatrix}
\sin\theta\cos\phi&
\sin\theta\sin\phi&
\cos\theta
\end{bmatrix}^{T},\\
\hat{\boldsymbol\theta}(\theta,\phi)
&=
\begin{bmatrix}
\cos\theta\cos\phi&
\cos\theta\sin\phi&
-\sin\theta
\end{bmatrix}^{T},\\
\hat{\boldsymbol\phi}(\phi)
&=
\begin{bmatrix}
-\sin\phi&
\cos\phi&
0
\end{bmatrix}^{T},\\
\mathbf U(\theta,\phi)
&=
\begin{bmatrix}
\hat{\boldsymbol\theta}(\theta,\phi)&
\hat{\boldsymbol\phi}(\phi)
\end{bmatrix}
\in\mathbb R^{3\times2}.
\end{aligned}
\label{eq:local_transverse_frame_conf}
\end{equation}

{For the known \ac{rhcp} sensing waveform, the Jones vector in the local transverse basis is defined as $\mathbf p_{\mathrm{RHCP}}=[-j,\ 1]^T/\sqrt2$. The Cartesian sensing-field phasor is therefore $\tilde{\mathbf E}_{\mathrm s}=\alpha\mathbf U(\theta,\phi) \mathbf p_{\mathrm{RHCP}}$, where $\alpha$ absorbs the unknown received \ac{rf}-field scale and common phase. The corresponding Rydberg steering vector is}
\begin{equation}
\mathbf a_{\mathrm R}(\theta,\phi) = \mathbf U(\theta,\phi)\mathbf p_{\mathrm{RHCP}} =
\frac{1}{\sqrt2}
\begin{bmatrix}
-j\cos\theta\cos\phi-\sin\phi\\
-j\cos\theta\sin\phi+\cos\phi\\
j\sin\theta
\end{bmatrix}.
\label{eq:rhcp_steering_conf}
\end{equation}
For the numerical link, a far-field free-space path sets the previously unspecified complex amplitude~\cite{itur_p525_5}:
\begin{equation}
\begin{aligned}
\alpha(r)&=A_{\mathrm s}(r)e^{j(\varphi_0-2\pi r/\lambda)},\\
A_{\mathrm s}(r)&=\sqrt{\frac{Z_0P_{\mathrm{tx}}G_{\mathrm{tx}}}{2\pi r^2}}
\simeq\frac{\sqrt{60P_{\mathrm{tx}}G_{\mathrm{tx}}}}{r}.
\end{aligned}
\label{eq:free_space_field_conf}
\end{equation}
Here $r$ is the link distance, $P_{\mathrm{tx}}$ is the transmit power, $G_{\mathrm{tx}}$ is the linear transmit gain towards the receiver, $Z_0$ is the free-space impedance, $\lambda=c/f_0$ is the carrier wavelength with $c$ the speed of light, and $\varphi_0$ is an unknown transmit phase. Since $\|\mathbf U\mathbf p_{\mathrm{RHCP}}\|=1$, $A_{\mathrm s}$ is the norm of the peak complex field. Both receivers observe the same incident field; the estimator still treats $\alpha$ as unknown.

After coherent beat extraction and response normalization, the sensing
observation is
\begin{equation}
\mathbf b_{\mathrm s}
=
\alpha\mathbf a_{\mathrm R}(\theta,\phi)
+
\mathbf w_{\mathrm s},
\label{eq:sensing_observation_conf}
\end{equation}
where $\mathbf w_{\mathrm s}\sim\mathcal{CN}(\mathbf 0,\mathbf R_w)$ is the
response-normalised Rydberg beat-domain noise.

\subsection{Maximum-Likelihood \acs{doa} Estimation}
\label{subsec:matched_likelihood_doa}

For a fixed direction $(\theta,\phi)$, the maximum-likelihood estimate of the unknown complex amplitude $\alpha$ is the covariance-weighted projection
\begin{equation}
\widehat{\alpha}(\theta,\phi)
=
\frac{
\mathbf a_{\mathrm R}^{H}(\theta,\phi)\mathbf R_w^{-1}\mathbf b_{\mathrm s}
}{
\mathbf a_{\mathrm R}^{H}(\theta,\phi)\mathbf R_w^{-1}\mathbf a_{\mathrm R}(\theta,\phi)
}.
\label{eq:alpha_hat_conf}
\end{equation}
Substituting \eqref{eq:alpha_hat_conf} into the Gaussian log-likelihood and dropping terms independent of $(\theta,\phi)$ gives the profiled likelihood score
\begin{equation}
S_{\mathrm R}(\theta,\phi)
=
\frac{
\left|
\mathbf a_{\mathrm R}^{H}(\theta,\phi)
\mathbf R_w^{-1}
\mathbf b_{\mathrm s}
\right|^2
}{
\mathbf a_{\mathrm R}^{H}(\theta,\phi)
\mathbf R_w^{-1}
\mathbf a_{\mathrm R}(\theta,\phi)
}.
\label{eq:matched_score_conf}
\end{equation}
The maximum-likelihood \ac{doa} estimate is $(\hat\theta,\hat\phi)=\arg\max_{(\theta,\phi)\in\mathcal S}
S_{\mathrm R}(\theta,\phi)$, where $\mathcal S$ is the angular search region. The denominator in \eqref{eq:matched_score_conf} removes sensitivity to the candidate steering vector norm, while the covariance weighting accounts for unequal or correlated Rydberg beat-domain noise channels.

\subsection{Coarse-to-Fine Angle Search}
\label{subsec:coarse_to_fine_search}
The resulting estimator is implemented by a coarse-to-fine angular search. The receiver first evaluates \eqref{eq:matched_score_conf} on a coarse grid $\mathcal G_{\mathrm c}$ over $\mathcal S$ and selects its peak $(\theta_{\mathrm c},\phi_{\mathrm c})$; it then repeats the same score evaluation on a local fine grid $\mathcal G_{\mathrm f}(\theta_{\mathrm c},\phi_{\mathrm c})$ around this peak to obtain $(\hat\theta,\hat\phi)$. The response estimate is obtained offline or updated infrequently, whereas the online sensing step extracts the current beat vector, normalises it with the stored $\widehat c_{\mathrm R}$, and performs this two-stage search.


\section{Numerical {Results}}
\label{sec:numerical_demonstration}
The numerical {simulation} compares the proposed Rydberg beat-vector receiver with a traditional \ac{va} benchmark. {The \ac{va} benchmark is an ideal calibrated tripole with an identity Cartesian response.} \bt{The setup uses $^{85}$Rb with the $5S_{1/2}$--$5P_{3/2}$--$60D_{5/2}$ \ac{eit} ladder and the $60D_{5/2}\leftrightarrow59F_{7/2}$ \ac{rf} transition at $f_0=10.663$ GHz, with $\|\mathbf d_{34}\|=1.7108\times10^{-26}$ C\,m. Each observation window has duration $T_{\mathrm{obs}}=T_i=0.2$ ms and is sampled at $f_{\mathrm s}=50$ MHz, giving $N_{\mathrm{obs}}=10^4$ samples; we further set $T_r=10~\mu$s and $N_a=5\times10^5$\cite{cui2025atomicmimo}. The \ac{lo} offsets are $(300,270,-200)$ kHz, with amplitude $20$ V/m per axis. The \ac{va} uses an antenna noise temperature of $T_{\mathrm{ant}}=290$ K.} Both receivers are evaluated at their fundamental noise limits: the \ac{qsn} floor for the Rydberg receiver and the antenna thermal noise for the \ac{va}.

The two receivers use the same external field-noise covariance, with $\sigma_{\mathrm{ext}}=1~\mu\mathrm{V/m}$ per axis for the observation. This fixed value represents additional environmental interference after extraction, rather than the thermal background already included in the \ac{va} antenna temperature. Independent noise realisations with the same external covariance are used in the two branches. The \ac{va} thermal noise is referred to the input electric field as
\begin{equation}
\begin{aligned}
\sigma_{E,\mathrm{th}}^2
&=\frac{2Z_0 k_{\mathrm B}T_{\mathrm{ant}}B_{\mathrm{eq}}}{A_{\mathrm{eff}}},\\
A_{\mathrm{eff}}&=\frac{G_{\mathrm{rx}}\lambda^2}{4\pi},\qquad
\mathbf R_{\mathrm{VA}}=(\sigma_{E,\mathrm{th}}^2+\sigma_{\mathrm{ext}}^2)\mathbf I_3.
\end{aligned}
\label{eq:va_field_noise_conf}
\end{equation}
Here $k_{\mathrm B}$ is the Boltzmann constant, $T_{\mathrm{ant}}$ is the antenna noise temperature, and $G_{\mathrm{rx}}=1.5$ is the nominal short-dipole gain used in the per-port field conversion, with unit radiation and matching efficiencies. The total two-sided complex-baseband noise bandwidth is $B_{\mathrm{eq}}=1/T_{\mathrm{obs}}=5$ kHz. The Rydberg branch uses the \ac{qsn} floor in \eqref{eq:qsn_field_floor_conf}; no additional bandwidth factor is applied to this time-integrated expression. The \ac{va} observation thus follows \eqref{eq:sensing_observation_conf} with noise covariance $\mathbf R_{\mathrm{VA}}$ and is processed by the same estimator.

Each trial uses a single $0.2$ ms observation window, without cross-window averaging. Both branches use the same coarse-to-fine maximum-likelihood search, with a $2^\circ$ coarse grid and a $0.05^\circ$ local fine grid spanning $\pm2^\circ$ around its peak. Accuracy is measured by the angular error, i.e., the great-circle separation between the true and estimated directions, $\epsilon_{\mathrm{ang}}=\cos^{-1}\!\big(\hat{\mathbf k}^{T}(\hat\theta,\hat\phi)\,\hat{\mathbf k}(\theta,\phi)\big)$, and the angular \ac{rmse} is the root-mean-square of $\epsilon_{\mathrm{ang}}$ over trials.

\subsection{Three-Axis Beat Extraction and \acs{doa} Sensing}

\begin{figure}[t]
\centering
\includegraphics[width=0.7\linewidth]{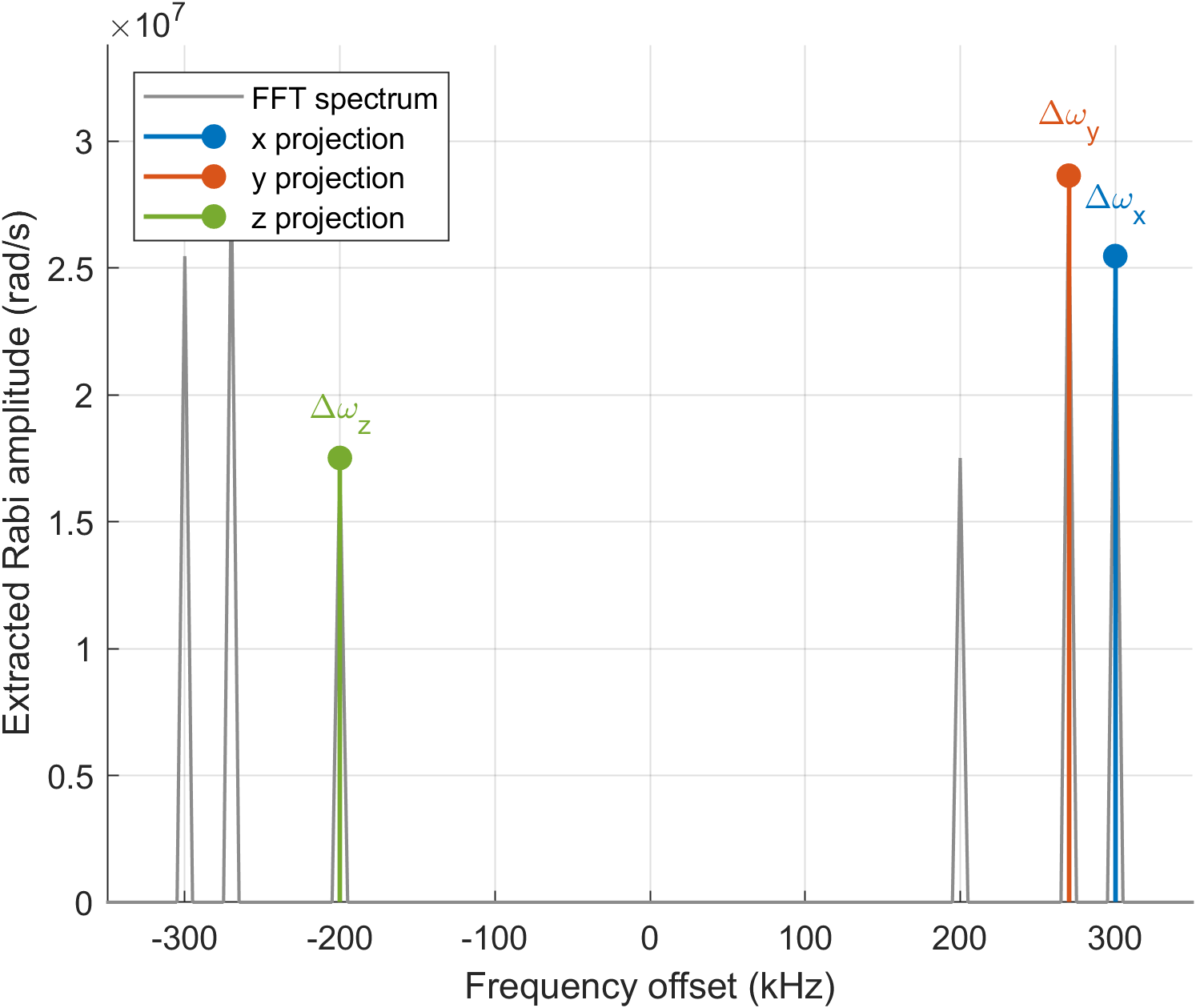}
\caption{\textbf{Three-axis beat extraction from a scalar Rabi-frequency trace.}
The representative \acs{rhcp} sensing field arrives from $(\theta,\phi)=(36^\circ,28^\circ)$. The three Cartesian projections are mapped to beat offsets $(\Delta f_x,\Delta f_y,\Delta f_z)=(300,270,-200)$ kHz and recovered from the scalar trace by \acs{fft} extraction.}
\label{fig:threefreq_beat_extraction}
\end{figure}

\bt{Fig.~\ref{fig:threefreq_beat_extraction}} verifies the front-end mechanism used in the equivalent beat-vector model. A single scalar Rabi-frequency trace $\Omega_{\mathrm{eff}}(t)$, i.e., the optical trace $y(t)$ with unit feature slope $s_{\Omega}$, is transformed to the frequency-offset domain, where the three retained \ac{fft} bins at $\Delta f_z=-200$ kHz, $\Delta f_y=270$ kHz, and $\Delta f_x=300$ kHz {represent} the Cartesian projections. Since the three axis gains are equalised, the differences between peak heights reflect the magnitudes of the \ac{rhcp} field projections onto the three axes for the chosen direction, while the frequency separation is imposed by the three-tone \ac{lo}. This result supports using the normalised vector $\mathbf b_{\mathrm s}$ as the three-axis observation passed to the \ac{doa} estimator in~\eqref{eq:sensing_observation_conf}.

\begin{figure}[t]
\centering
\includegraphics[width=0.7\linewidth]{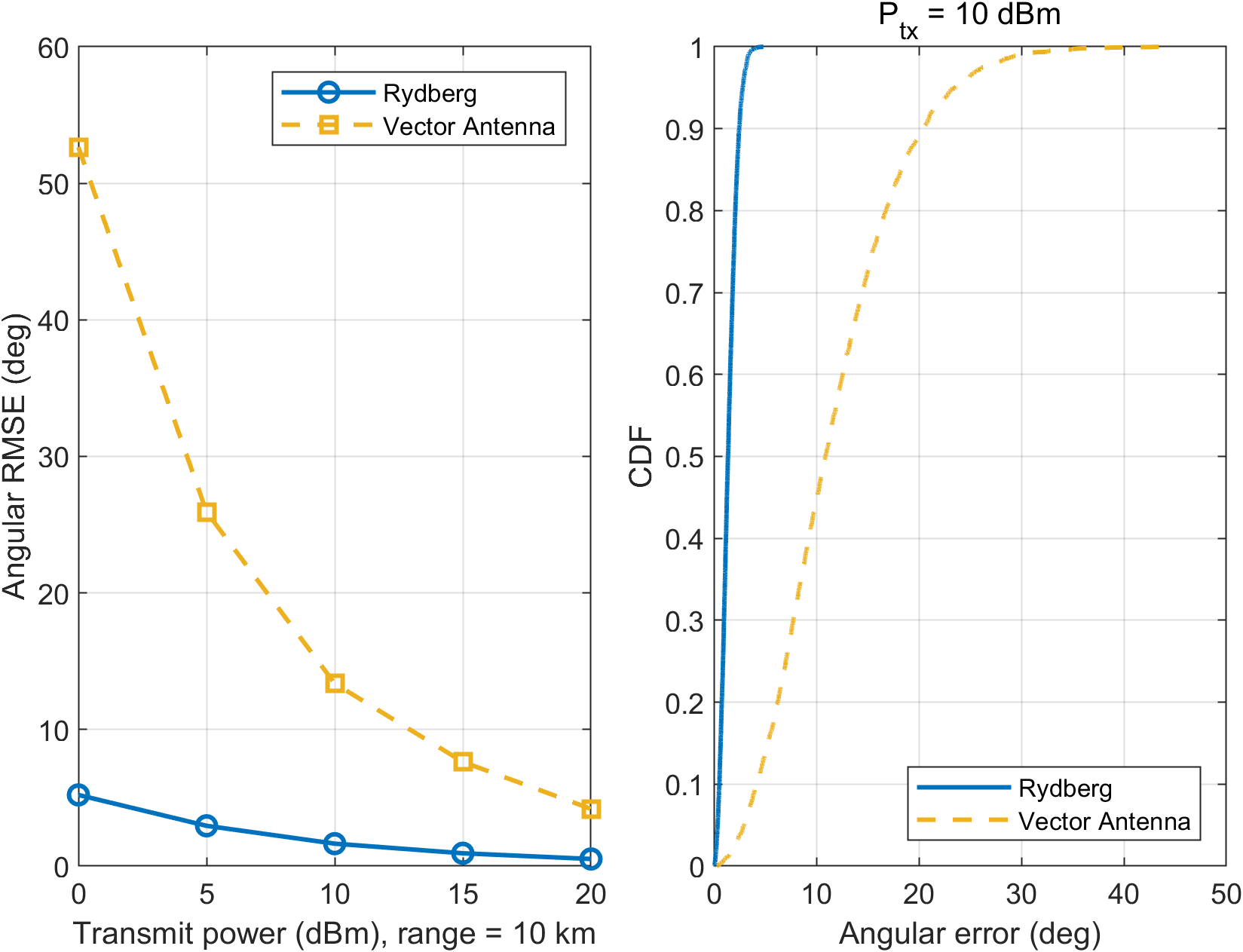}
\caption{\textbf{Random-\acs{doa} comparison.} Angular \acs{rmse} versus $P_{\mathrm{tx}}$ in dBm and angular-error \acs{cdf} at $10$ dBm, for $r=10$ km and $G_{\mathrm{tx}}=0$ dBi. Both receivers use the same coarse-to-fine angular search.}
\label{fig:doa_rmse_vs_tx_power}
\end{figure}

\bt{Fig.~\ref{fig:doa_rmse_vs_tx_power}} shows the angular \ac{rmse} versus absolute transmit power. At each level, 3000 independent trials sample $\phi$ uniformly over $[-180^\circ,180^\circ]$ and $\cos\theta$ uniformly over $[-1,1]$. Because the extraction in \eqref{eq:raw_beat_extraction_conf} is linear and the three tones are bin-centred, the extracted and normalised beat vector is statistically equivalent to \eqref{eq:sensing_observation_conf}; the Monte Carlo trials therefore draw $\mathbf b_{\mathrm s}$ directly with the covariances specified above, while Fig.~\ref{fig:threefreq_beat_extraction} verifies the extraction step itself. The angular-error \ac{cdf} is evaluated at $P_{\mathrm{tx}}=10$ dBm. The Rydberg angular \ac{rmse} decreases from $5.21^\circ$ at $0$ dBm to $1.64^\circ$ at $10$ dBm and $0.52^\circ$ at $20$ dBm, compared with $52.65^\circ$, $13.37^\circ$, and $4.19^\circ$ for the \ac{va}. At $10$ dBm, the median/90th/95th percentiles are $1.38^\circ$/$2.47^\circ$/$2.83^\circ$ for Rydberg and $10.90^\circ$/$20.41^\circ$/$23.55^\circ$ for the \ac{va}.

\begin{figure}[t]
\centering
\includegraphics[width=0.7\linewidth]{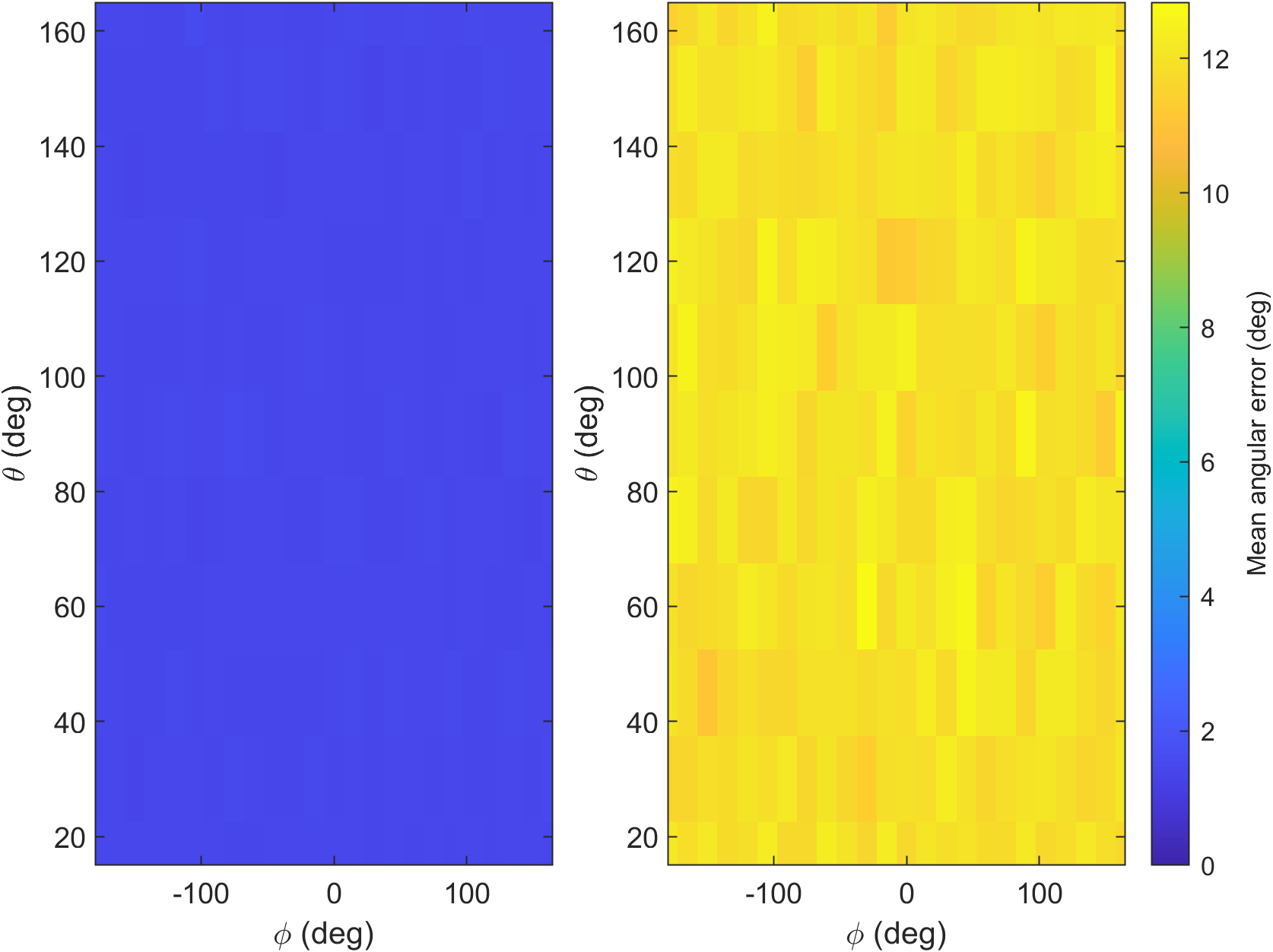}
\caption{\textbf{Angular-error heatmaps at $P_{\mathrm{tx}}=10$ dBm.} The true-direction grid is $\theta=15^\circ:15^\circ:165^\circ$ and $\phi=-180^\circ:15^\circ:165^\circ$, with 500 noise realisations per point. The two panels use the same colour scale and coarse-to-fine estimator.}
\label{fig:random_doa_error_heatmap}
\end{figure}

\bt{Fig.~\ref{fig:random_doa_error_heatmap}} shows the spatial error distribution at $P_{\mathrm{tx}}=10$ dBm. Each true direction averages 500 independent noise realisations. The mean angular error across the displayed directions is $1.46^\circ$ for Rydberg and $11.97^\circ$ for the \ac{va}. Both maps are approximately direction-uniform, as expected for ideal calibrated Cartesian responses with isotropic noise.

\section{Conclusion}
\label{sec:conclusion}

This paper formulated a three-axis Rydberg beat-vector readout as a response-normalised vector-field sensor for {\ac{rf}} \ac{doa} estimation. Distinct \ac{rf} \ac{lo} offsets separate the Cartesian field projections into resolvable optical beat tones, while response calibration converts the extracted {beat} coefficients into a three-axis sensing vector. For a known circularly polarised incident \ac{rf} field, the proposed Rydberg receiver induces a direction-dependent steering vector and enables maximum-likelihood \ac{doa} estimation with unknown field amplitude. The numerical results verify the complete sensing chain from scalar trace to beat extraction and angular search. The free-space link model provides a common incident-field basis for comparison with the traditional \ac{va} benchmark; at $10$ dBm and $10$ km, the angular \ac{rmse} is $1.64^\circ$ for the ideal \ac{qsn}-limited Rydberg receiver with external noise and $13.37^\circ$ for the calibrated \ac{va}. The current work focuses on the principle of polarization-based \ac{doa} of Rydberg \ac{rf} receiver. We will explore mechanisms for handling multipath, polarisation-distorting channels, and investigate how vapor cell dimensions affect the angular response, while simulating broader scenarios that vary distance, transmit gain, and residual calibration errors. 

\bibliographystyle{IEEEtran}
\bibliography{IEEEabrv,Bibliography,IEEE_ISAC_link}

\end{document}